\documentclass[final]{elsart}
\usepackage{graphicx}
\usepackage{amssymb,amsmath,url}
\newcommand{\mat}[1]{\mathbf{#1}}
\newcommand{\vecr}{\mathbf{r}}
\newcommand{\vecrp}{\mathbf{r'}}

\begin{document}

\begin{frontmatter}
\title{Approximation formulae for the acoustic radiation impedance of a cylindrical pipe}
\author[LMA]{F. Silva\corauthref{cor}}
\corauth[cor]{Corresponding author.}
\ead{silva@lma.cnrs-mrs.fr}
\author[LMA]{Ph. Guillemain}
\ead{guillemain@lma.cnrs-mrs.fr}
\author[LMA]{J. Kergomard}
\ead{kergomard@lma.cnrs-mrs.fr}
\author[LMA]{B. Mallaroni}
\ead{mallaroni@lma.cnrs-mrs.fr}
\author[RUTGERS]{A. N. Norris}
\ead{norris@rutgers.edu}

\address[LMA]{Laboratoire de M\'ecanique et d'Acoustique, UPR CNRS 7051,
    31 chemin Joseph Aiguier, 13402 Marseille cedex 20, France}
\address[RUTGERS]{Mechanical \& Aerospace Engineering, Rutgers University,
    Piscataway NJ 08854-8058, USA }

\begin{abstract}
Useful approximation formulae for radiation impedance are given for the reflection coefficients of both infinitely flanged and unflanged rigid-walled cylindrical ducts. The expressions guarantee that simple but necessary physical and mathematical principles are met, like hermitian symmetry for the reflection coefficient (identical behaviour of positive and negative frequencies) and causality for the impulse response. A non causal but more accurate expression is also proposed that is suitable for frequency-domain applications. The formulae are obtained by analytical and numerical fitting to reference results from Levine \& Schwinger for the unflanged case and extracted from the radiation impedance matrix given by Zorumski for the infinite flanged case.
\end{abstract}

\end{frontmatter}

\section{Introduction and results}
\label{sec:intro}
The problem of the radiation acoustic impedance of the planar mode in a circular pipe with rigid walls (i.e. with homogeneous Neumann boundary condition) is a classical problem of acoustics. Various detailed calculations for both unflanged and infinitely flanged ducts have already been provided, see e.g. \cite{levine:1948,nomura:1960,ando:1969,norris:1989}. In addition, experimental investigations with various flanges have been compared to theoretical and numerical results \cite{dalmont:2001c}. There is a need for approximated formulae, such as those given by \cite{norris:1989,davies:1980b} or \cite{causse:1984}. Unfortunately these formulae do not fulfil the conditions for a physically representative model, as for instance the hermitian property of the reflection coefficient (see e.g. \cite{rienstra:2006}):
\begin{equation}
\mathcal{R}(-\omega )=\mathcal{R}^*(\omega )
\label{eq:hermitian}
\end{equation}
where $*$ means "complex conjugate of", or the causality of the impulse response of the reflection coefficient, obtained by inverse Fourier transform and denoted $r(t)=\text{FT}^{-1}(\mathcal{R}(\omega))$, also known as the reflection function. This is essential to ensure that the time domain signals are real causal quantities.
\par The aim of this short paper is to provide suitable approximate formulae based on accurate analytical or numerical fitting. Choosing the end of the bore as the reference plane and assuming a time dependence $\exp{}(-j\omega t)$, the principal results will be expressed in terms of the pressure reflection coefficient $\mathcal{R}(\omega)$:
\begin{equation}
\mathcal{R} = -|\mathcal{R}|e^{2jkL}
=\frac{\mathcal{Z}_{r}-1}{\mathcal{Z}_{r}+1}.
\label{eq:ZR}
\end{equation}
The dimensionless (i.e. divided by the characteristic impedance $\rho c/\pi a^2$) radiation impedance $\mathcal{Z}_{r}$ can then be expressed in terms of $\mathcal{R}$ as follows:
\begin{equation}
\mathcal{Z}_{r}=\frac{1+\mathcal{R}}{1-\mathcal{R}}
=-j\tan\left(kL-j\frac{1}{2}\ln{|\mathcal{R}|}\right).
\label{eq:ZR2}
\end{equation}
$k=\omega/c$ and $L$ denote the acoustic wave number and the end correction due to radiation, respectively.
\par Section~\ref{sec:ref} describes how reference values were calculated. The requirements are discussed in Section~\ref{sec:model}. The approximate formulae are presented together with numerical results in Section~\ref{sec:results}. Table~\ref{tab:parameters} summarises the various formulae in the frequency and time domains.

\section{Calculation of the reference values}
\label{sec:ref}
\subsection{Unflanged case}
Air vibrations in a rigid cylindrical pipe of negligible wall thickness and radius $a$ propagate into free space through a circular sharp-edged opening. Nonlinear effects like shock wave, vortex shedding, or mean axial flow are not considered. Using the Wiener-Hopf technique, Levine \& Schwinger\cite{levine:1948} obtained an integral formulation of the reflection coefficient $\mathcal{R}(ka)$ for the planar mode, for frequencies below the cutoff frequency of the first non planar axisymmetric mode. We performed numerical evaluations of Eq.~(V.16) of Ref.~\cite{levine:1948} using the \texttt{quadv} function from \texttt{Matlab}\cite{matlab}. The values obtained were compared to the asymptotic approximations given by Eqs.~(VII.1) and (VII.2) of Ref.~\cite{levine:1948}, with a maximum deviation compatible with the value of $3\%$ mentioned in that paper.

\subsection{Flanged case}
The reflection coefficient for an infinitely flanged pipe has been calculated by Nomura \emph{et al}\cite{nomura:1960} using Weber-Schafheitlin integrals and by Norris \& Sheng\cite{norris:1989} using modal expansion of the duct pressure field and a Green's function representation. The latter results in a modal sum for the planar mode reflection coefficient, each coefficient having a complex integral expression. 
\par It is also possible to derive the radiation impedance for the planar mode from the radiation impedance matrix written as in Ref.~\cite{zorumski:1973}, considering only the modes being symmetrical about the axis of the pipe. The pressure and the velocity fields at the end of the bore are expressed in terms of the duct modes:
\begin{equation}
P(\vecr,k)=\rho c^2\sum_{n\geq 0} P_n \psi_n(\vecr,k) 
\quad \text{ and }\quad
V(\vecr,k)=c \sum_{n\geq 0} V_n \psi_n(\vecr,k),
\end{equation}
with $\rho$ the mean air density, and $c$ the sound velocity in the free space, respectively. The dimensionless coefficients $P_n$ and $V_n$ are linked by $\mat{P}=\mat{Z}\mat{V}$ with:
\begin{equation}
\mat{Z}_{n,m} = \frac{(jk)^3}{2\pi}
\iint_{\vecr,\vecrp}
\frac{e^{jk|\vecr-\vecrp|}}{2\pi |\vecr-\vecrp|}
\psi_n(\vecr,k)\psi_m(\vecrp,k) dS(\vecr)dS(\vecrp)
\quad \forall n,\,m\in [0,N],
\end{equation}
where the $\psi_n(\vecr)$ are the normalised Bessel basis functions in the duct as in Eq.~(8) of Ref.\cite{zorumski:1973}:
\begin{equation}
\psi_n(\vecr,k)=\frac{\sqrt{2}}{ka}\frac{J_0(j_n r/a)}{|J_0(j_n)|},
\end{equation}
$j_n$ being the $n$th zero of the Bessel function $J_1$.
Computation of the impedance matrix has been performed using Eq.~(24) of Ref.~\cite{zorumski:1973}, giving results equal to the ones given by formulation given by Eq.~(13) of Ref.~\cite{norris:1989}.
\par
\begin{figure}[tbp]
\begin{center}
\includegraphics[width=0.9\columnwidth]{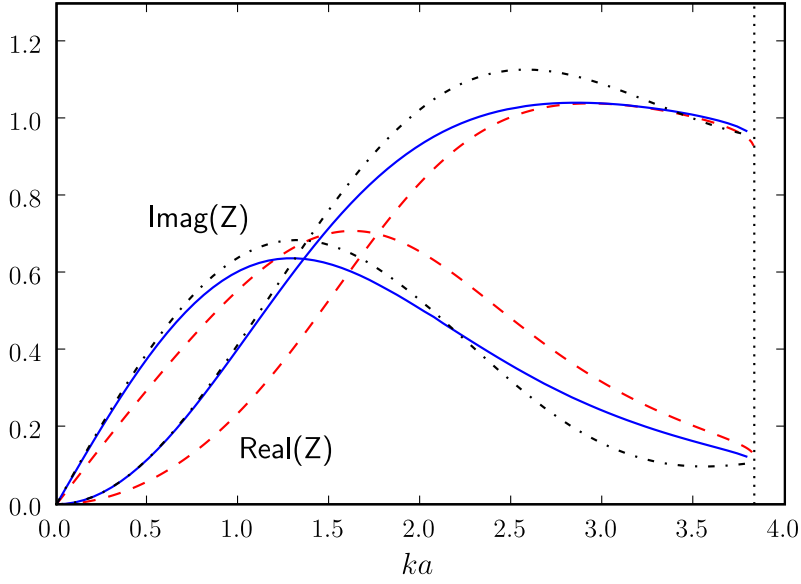}
\caption{Real and imaginary parts of the radiation impedances of the circular piston set in an infinite plane baffle ($z_{00}$, dash-dotted curves) and of the planar mode of a cylinder in the flanged ($\mathcal{Z}_r$, solid curves) and unflanged cases (dashed lines).}
\label{fig:Z}
\end{center}
\end{figure}

 The vectors $\mat{P}$ and $\mat{V}$ and the matrix $\mat{Z}$ are then decomposed into blocks, separating planar and non-planar evanescent components of the pressure and the flow fields:
\begin{equation}
\begin{bmatrix} P_{0}\\ \mat{P'}\end{bmatrix}
=\begin{bmatrix} z_{00} & \mat{z}^T\\\mat{z} & \mat{Z'}\end{bmatrix}
 \begin{bmatrix}V_{0}\\ \mat{V'}\end{bmatrix},
\end{equation}
where $z_{00}$ is the dimensionless radiation impedance of a circular piston and $\mat{z}$ gives the coefficients of the higher-order modes for an incident plane velocity field $\mat{V}(\vecr)=V_0$. Provided that the non-planar modes do not propagate, i.e. $ka<j_1\simeq 3.832$, and that the upstream bore is sufficiently long so that the evanescent modes do not meet a reflective obstacle in the duct, the following relation links the non-planar components:
\begin{equation}
\mat{P'}=-\mat{Z'}_{c} \mat{V'}
\end{equation}
where $\mat{Z'}_{c}$ is the diagonal matrix of the dimensionless characteristic impedances $\mat{Z'}_{c,n}=k/k_n$, with $k_n a=j\sqrt{j_n^2-(ka)^2}$ the wave number of evanescent modes. Eventually, planar components of pressure and flow are such that:
\begin{gather}
P_0 = z_{00} V_0 + \mat{z}^T \mat{V'}
= \left(z_{00}-\mat{z}^T(\mat{Z'}+\mat{Z'}_{c})^{-1}\mat{z}\right) V_0 =\mathcal{Z}_r V_0
\end{gather}
this expression showing the influence of evanescent non-planar modes present at the opening on the effective plane radiation impedance. The reflection coefficient for the planar mode is then obtained by means of Eq.~(\ref{eq:ZR}). Fig.~\ref{fig:Z} shows the effects of the flange (dashed vs solid lines) and of the production of higher components at the end of the bore (solid vs dash-dotted lines).
\par The numerical method described here has been compared to the first coefficient $R_{000}$ of the generalised reflection coefficient given by Zorumski (Eq.~(32) of Ref.~\cite{zorumski:1973}) and to the method given by Norris (Eqs.(11) and (14) of Ref.\cite{norris:1989}) where the $\alpha_{nm}$ coefficients can be deduced from the radiation impedance matrix. Results are in very good agreement, with relative error less than $10^{-6}$. Convergence of the calculations with respect to the number of higher order modes taken into account appeared to be significant for 10 modes. Matrix calculations were done using \texttt{LAPACK}\cite{lapack}, a standard linear algebra library.

\section{Model requirements}
\label{sec:model}
Norris \& Sheng (see Eq.~(27) of Ref.~\cite{norris:1989}) attempted to provide simple accurate formulae for $|\mathcal{R}|$ with a rational function. But a non negligible drawback of these formulae (and of other low-frequency approximations, see e.g. \cite{davies:1980b,causse:1984}) is that they are not hermitian, i.e. the property $\mathcal{R}(- \omega)=\mathcal{R}(\omega)^*$ is not satisfied, and cannot be used for both positive and negative frequency domains. Another point is that it was not possible to estimate a time-domain reflection function $r(t)$ because the approximated formulae were not applicable for higher frequencies (the expression for the modulus $|\mathcal{R}(\omega)|$ become negative when $\omega$ increases). In this section the complete set of requirements on $\mathcal{R}(\omega)$ is addressed.

\subsection{Low frequency behaviour of the reflection coefficient}
The first requirement on the reflection coefficient is satisfaction of the following asymptotic forms in the low frequency domain:
\begin{equation}
|\mathcal{R}|(\omega\rightarrow 0)
=1-\beta (ka)^2+o(ka)^2 \mbox{ with } \beta=
\begin{cases}
1/2 &\text{for the unflanged case},\\
1 &\text{for the flanged case},
\end{cases}
\end{equation}
using the \emph{little-o notation}. In the case of the baffled circular piston, the truncated part of the expansion is known to grow as $(ka)^4$, but is more complicated in the unflanged case. Concerning the length correction as defined in Eq.~(\ref{eq:ZR}), it becomes:
\begin{equation}
\frac{L}{a}(\omega\rightarrow 0)=\eta=
\begin{cases}
0.6133 &\text{for the unflanged case},\\
0.8216 &\text{for the flanged case},
\end{cases}
\end{equation}
these static values being the ones given by \cite{levine:1948} and \cite{norris:1989}.

\subsection{Impulse response}
In order to enable the use of a time-domain reflection function, the condition for inverse Fourier transform existence has to be fulfilled. First of all, $|\mathcal{R}(\omega)|$ has to tend to zero at high frequencies. The impulse response $r(t)$ must also be a real quantity, which is guaranteed by the hermitian property in the frequency-domain (see e.g Eq.~\ref{eq:hermitian}).
\par Another physics-driven requirement is that the impulse response should be causal. The Kramers-Kronig relations\cite{mcdaniel:1999} provide a means to express this requirement in the frequency-domain. Another method is to study the placement of the poles of the reflection coefficient. Using the time dependence $\exp{(-j\omega t)}$, the poles have to be located in the complex half-plane $\mathrm{Im}(\omega)<0$ (see e.g. \cite{rienstra:2006}, with the opposite time dependence convention); otherwise anticausal components appear in the inverse Fourier transform. However, it is not possible to calculate a time-domain impulse response based on the equations described in the previous section as this computation requires the knowledge of $\mathcal{R}(\omega)$ on the full frequency range. As a practical alternative, we aim to provide approximations having an extended domain of validity in both frequency and time.
\par We notice that if the reflection coefficient is causal, the input impedance is causal as well, because the modulus of $\mathcal{R}(\omega)$ is less than unity. Expanding $(1-\mathcal{R})^{-1}$ in Eq.~(\ref{eq:ZR2}) in the form of $1+\mathcal{R}+\mathcal{R}^2+\ldots$, the inverse Fourier transform of $\mathcal{Z}_r$ is expressed as an infinite series of terms involving convolution products of causal functions, thus the result is a causal function. A physical interpretation in terms of successive reflections is classical. This reasoning can not be done in a reciprocal manner using Eq.~(\ref{eq:ZR}), since the modulus of the dimensionless impedance can be large.

\section{Approximate formulae and results}
\label{sec:results}
Two models are presented in this section. Both models satisfy the requirements mentioned in Section~\ref{sec:model}: they are causal in the time domain and satisfy the hermitian property and the desired low-frequency behaviour in the frequency domain. In particular they both result in very similar accuracy in the frequency domain below the below cutoff frequency $ka=j_1$. The main difference is in the high-frequency behaviour which results in different smoothness properties at $t=0$.

\subsection{Model ($\nu$, $\alpha$)}
The impulse response is modelled by the following expression:
\begin{equation}
r(t)=-A\left(\frac{ct}{a}\right)^\nu 
\exp{(-\alpha \frac{ct}{a})}
\text{ for $t>0$, $0$ otherwise.}
\label{eq:modelnu}
\end{equation}
This definition ensures causality and hermitian symmetry (with real $\alpha$ and $\nu$). Its frequency domain representation (with the convention $\exp{-j\omega t}$) is given by:
\begin{equation}
\mathcal{R}(\omega)
=-\frac{A a\Gamma(\nu+1)}{c\alpha^{\nu+1}} \left(1-\frac{jka}{\alpha}\right)^{-(\nu+1)}.
\end{equation}
\par
\begin{figure}[tbp]
\begin{center}
\includegraphics[width=0.9\columnwidth]{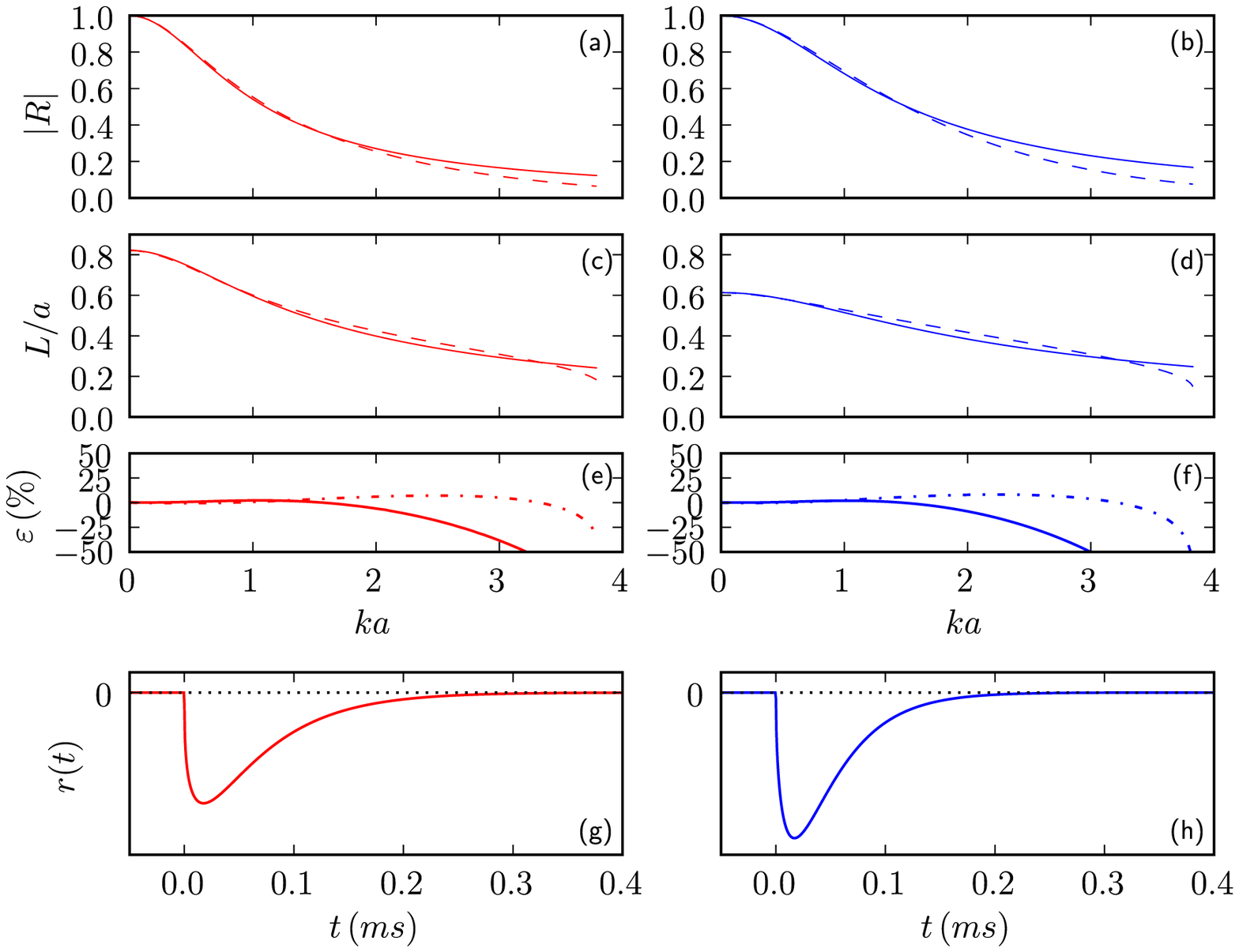}
\caption{Comparison in frequency domain of the references (dashed lines) and model ($\nu$,$\alpha$) of Eqs.~(\ref{eq:modelnu}) through (\ref{eq:modelnu_fin}) (in solid lines) for the unflanged (left column) and infinitely flanged (right column) cases. (a,b): modulus $|\mathcal{R}(ka)|$ of the reflection coefficient (approximation errors in solid lines in (e,f)). (c,d): dimensionless length correction $L/a$ (approximation errors in dash-dotted lines in (e,f)). (g,h): example of the reflection function ($a=7\mbox{mm}$).}
\label{fig:modele_nu}
\end{center}
\end{figure}

 The adjustment of the low-frequency behaviour allows the determination of the parameters $A$, $\nu$ and $\alpha$:
\begin{equation}
A = \frac{c\alpha^{\nu+1}}{a\Gamma(\nu+1)},\qquad
\alpha = \frac{\eta}{\beta}\qquad
\text{and}\qquad
\nu+1 = \frac{2\eta^2}{\beta}
\end{equation}
The value of $A$ leads to the following expression~:
\begin{equation}
\mathcal{R}(\omega)=-\left(1-\frac{jka}{\alpha}\right)^{-(\nu+1)}.
\label{eq:modelnu_fin}
\end{equation}
Numerical values of the parameters are given in Table~\ref{tab:parameters} and frequency and time responses are displayed in Fig.~\ref{fig:modele_nu} showing good agreement in frequency domain up to $ka=2$ and monotonic decrease above. The time-domain reflection function as defined by  Eq.~(\ref{eq:modelnu}) is causal, as expected.

\subsection{Low order Pad\'e approximant}
In order to permit analytical simplicity and possible calculations in the frequency domain, it is convenient to assume that the coefficient reflection depends on integers powers of frequency. That is not the case for the model ($\nu$,$\alpha$) which has a non-integer exponent $\nu$.
\par
\begin{figure}[tbp]
\begin{center}
\includegraphics[width=0.9\columnwidth]{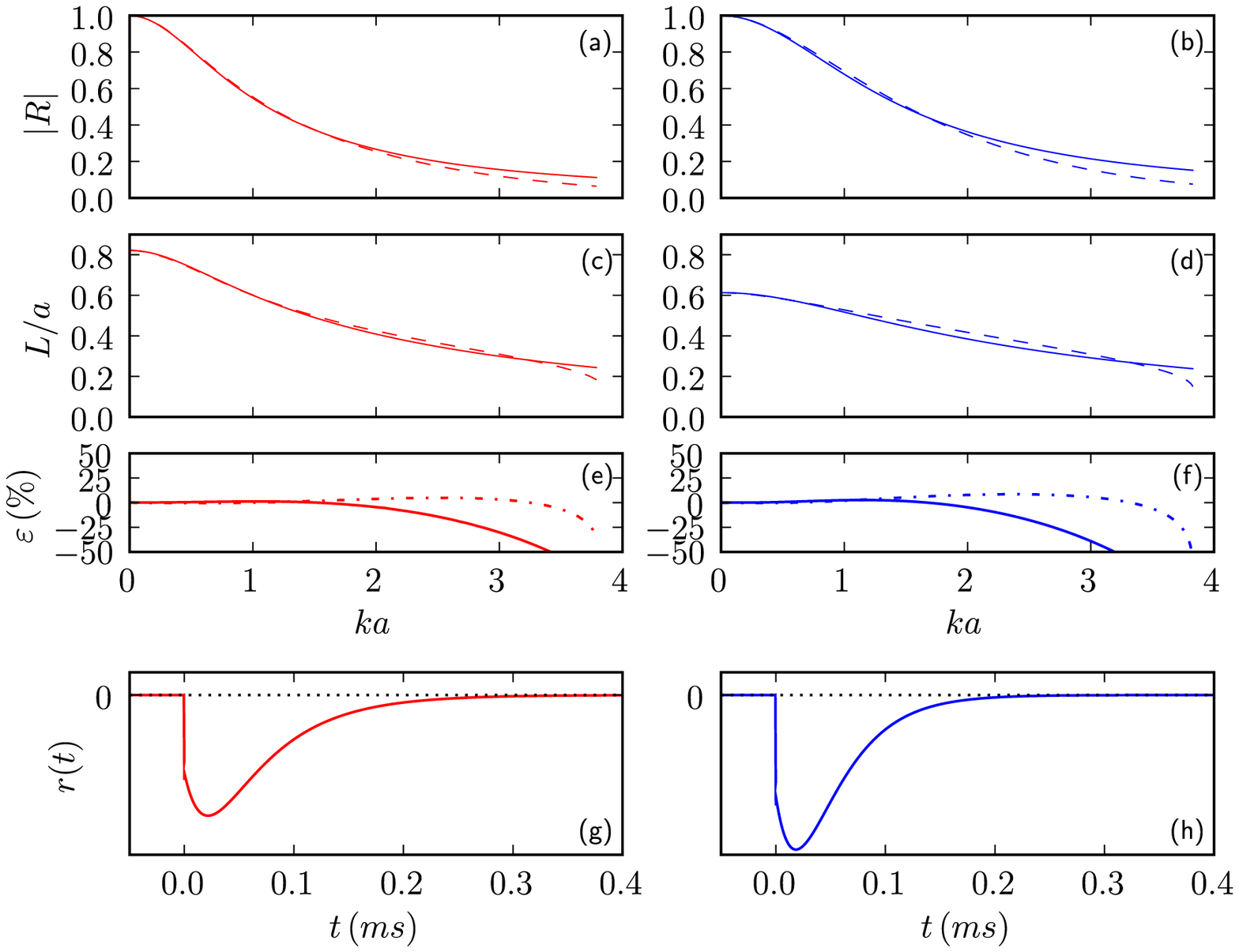}
\caption{Comparison in frequency domain of the references (dashed lines) and model (1,2) of Eqs.~(\ref{eq:model12}) through (\ref{eq:model12_fin}) (in solid lines) for the unflanged (left column) and infinitely flanged (right column) cases. (a,b): modulus $|\mathcal{R}(ka)|$ of the reflection coefficient (approximation errors in solid lines in (e,f)). (c,d): dimensionless length correction $L/a$ (approximation errors in dash-dotted lines in (e,f)). (g,h): example of the reflection function ($a=7\mbox{mm}$).}
\label{fig:modele_12}
\end{center}
\end{figure}

 A rational function approximation of the form:
\begin{equation}
\mathcal{R}(\omega)=-\frac{1-n_1 jka}{1-d_1 jka+d_2 (jka)^2}
\label{eq:model12}
\end{equation}
has been fitted to reference values. Low-frequency behaviour constrains the parameters $n_1$, $d_1$ and $d_2$ according to the following relations:
\begin{equation}
d_1-n_1 = 2\eta,
\qquad\mbox{ and }\qquad
d_1^2-n_1^2-2d_2 = 2\beta,
\end{equation}
so that a one-dimensional fitting is done using a Nelder-Mead simplex algorithm to minimise the error between the complex evaluations of the reference and the model. $N=75$ values of $\mathcal{R}(ka)$ were computed for regularly spaced values of $ka$ between $0$ and $j_1$ (the cutoff value for the first axisymmetric non planar mode). Numerical values obtained by constrained optimisation (requiring causality or, equivalently, that no pole lie in $\mathrm{Im}(\omega)>0$) are given in Table~\ref{tab:parameters}. In the same manner as for the previous model, frequency and time responses are shown in Fig.~\ref{fig:modele_12}. $\mathcal{R}(\omega)$ seems almost as well adjusted to the reference values as the model ($\nu$,$\alpha$) but, unlike that model, the impulse response $r(t)$ now shows an instantaneous initial step at $t=0$:
\begin{equation}
r(t>0) = \frac{c}{ad_2(\gamma_1-\gamma_2)}
\left(
 (n_1\gamma_2-1)e^{-\gamma_2\frac{ct}{a}}
-(n_1\gamma_1-1)e^{-\gamma_1\frac{ct}{a}}
\right),
\label{eq:rt12}
\end{equation}
where
\begin{equation}
\gamma_{1,2} = \frac{1}{2d_2}\left(d_1\pm\sqrt{d_1^2-4d_2}\right).
\label{eq:poles12}
\end{equation}
\par The radiation impedance $\mathcal{Z}_r$ is given by the following expression:
\begin{equation}
\mathcal{Z}_r(\omega)=\frac{(d_1-n_1)jka-d_2(jka)^2}{2-(d_1+n_1)jka+d_2(jka)^2}
\label{eq:model12_fin}
\end{equation}
similar to the one suggested by Doutaut \& Chaigne\cite{doutaut:1998}, where the coefficients were obtained by numerical fitting on the approximated radiation impedance given by Causs\'e \emph{et al.}\cite{causse:1984}.

\section{Relaxing the causality constraint}
The formulae given previously may appear somewhat inaccurate especially as the cutoff frequency of the first higher order mode is approached. However, it should be kept in mind that they are intended to simultaneously approximate the low frequency behaviour of the reflection coefficient, satisfy the hermitian property, and produce a causal physical response. With all these constraints being satisfied, the relative errors on the modulus and the length correction are less than $8\%$ for $ka\leq 2$. For applications where the time-domain response is not a critical criteria, relaxation of the causality constraint may lead to improved approximations.
\par
\begin{figure}[tbp]
\begin{center}
\includegraphics[width=0.9\columnwidth]{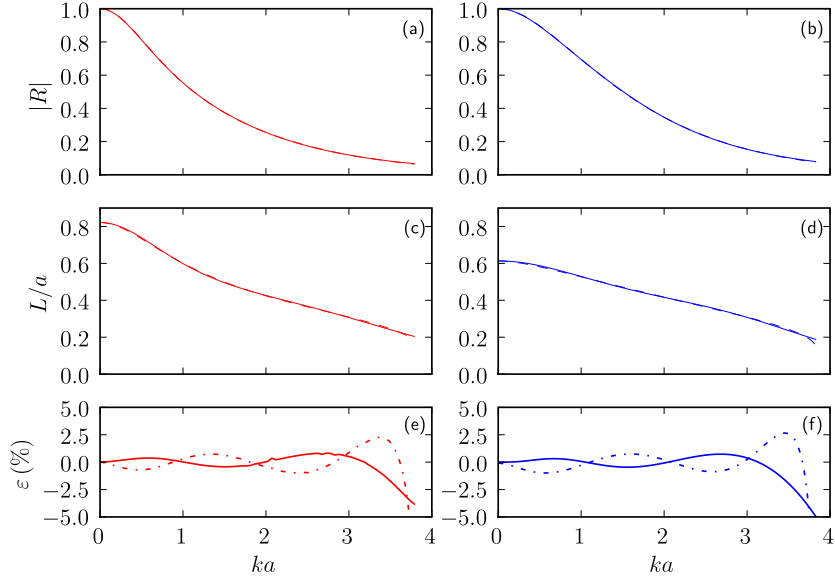}
\caption{Comparison in frequency domain of the references (dashed lines) and the non causal model of Eqs.~(\ref{eq:nc_module}) and (\ref{eq:nc_argument}) (in solid lines) for the unflanged (left column) and infinitely flanged (right column) cases. (a,b): modulus $|\mathcal{R}(ka)|$ of the reflection coefficient (approximation errors in solid lines in (e,f)). (c,d): dimensionless length correction $L/a$ (approximation errors in dash-dotted lines in (e,f)).}
\label{fig:modele_NC}
\end{center}
\end{figure}

 We find that modelling $|\mathcal{R}(ka)|$ and $L(ka)/a$ with Pad\'e approximants of order $(2,6)$ provides the more accurate approximations with the following expressions:
 \begin{gather}
|\mathcal{R}|=\frac{1+a_1(ka)^2}{1+(\beta+a_1)(ka)^2+a_2(ka)^4+a_3 (ka)^6},
\label{eq:nc_module}\\
\frac{L}{a}= \eta\frac{1+b_1(ka)^2}{1+b_2(ka)^2+b_3(ka)^4+b_4(ka)^6}
\label{eq:nc_argument},
\end{gather}
and the numerical values given in Table~\ref{tab:parameters}. These formulae fulfil all of the requirements except the causality principle, and approximation errors for the modulus and the length correction are less than $2\%$ for $ka<3$ for both unflanged and infinitely flanged cases as shown in Fig.~\ref{fig:modele_NC}. Combining formulae~(\ref{eq:nc_module}) and (\ref{eq:nc_argument}) into Eq.~(\ref{eq:ZR2}) allows the computation of the acoustic radiation impedance.

\section{Conclusion}
Practical approximation formulae of the acoustic radiation impedance of tubes are required for a variety of applications in acoustics. The necessity for such useful formulae is illustrated, in part, by the list of references cited here, which is long but certainly not exhaustive. The present results attempt to satisfy all the criteria required of the realistic physical system, such as causality, hermitian response and faithful low frequency behaviour. We therefore believe that the present contribution can be useful for calculations in the both time and frequency domains. Practical applications include, for instance, measurements in duct acoustics, musical acoustics, or loudspeakers enclosures where mean flow is absent or very slow.
\par Difficulty remains for the digital (sampled time) domain. In fact, the approximate formulae provided in this paper are all expressed in terms of the variable $ka$. A digital filter modelling the radiation at the end of a cylindrical duct has to be written as a function of the $z=\exp{j\omega T_e}$ variable (in the $T_e$-sampled time domain) and the coefficient of the filter are then dependent on the radius $a$. This could present difficulties if one wants to obtain approximation formulae suitable for the digital domain.
\par Despite such remaining technical issues, the calibrated formulae summarised in Table~\ref{tab:parameters} offer the acoustician practical and useful formulae for realistic applications. The numerical results of the calculations described in Section~\ref{sec:ref} are available at the following url: \url{http://www.lma.cnrs-mrs.fr/~PIM/DATA/}.
\begin{table}
\begin{center}
\begin{tabular}{|c||c|c|}
\hline
Model & Unflanged case & Flanged case\tabularnewline
& $\beta=1/2,\,\eta=0.6133$ & $\beta=1,\,\eta=0.8216$\tabularnewline 
\hline\hline
$\begin{array}{c}
    \mathcal{R}(\omega)=-\left(1-\frac{jka}{\alpha}\right)^{-(\nu+1)}\\
    r(t) = -A\left(\frac{ct}{a}\right)^\nu \exp{(-\alpha \frac{ct}{a})}\end{array}$
& $\begin{array}{c}\alpha=1.2266\\ \nu=0.504\\
    A = 1.534 \left.\frac{c}{a}\right. \end{array}$
& $\begin{array}{c}\alpha=0.8216\\ \nu=0.350\\
    A = 0.861 \left.\frac{c}{a}\right.\end{array}$
\tabularnewline\hline
$\mathcal{R}(\omega)=-\frac{1-n_1 jka}{1-d_1 jka+d_2 (jka)^2}$
& $\begin{array}{c}n_1 = 0.167 \\ d_1=1.393, \\ d_2=0.457\end{array}$
& $\begin{array}{c}n_1 = 0.182 \\ d_1=1.825, \\ d_2=0.649\end{array}$
\tabularnewline\hline\hline
$\begin{array}{c}
|\mathcal{R}|=\frac{1+a_1(ka)^2}{1+(\beta+a_1)(ka)^2+a_2(ka)^4+a_3 (ka)^6},\\
\frac{L}{a}= \eta\frac{1+b_1(ka)^2}{1+b_2(ka)^2+b_3(ka)^4+b_4(ka)^6},\\
\mbox{non causal}\end{array}$
& $\begin{array}{c} a_1=0.800\\ a_2=0.266\\ a_3=0.0263\\
    b_1=0.0599\\ b_2=0.238\\ b_3=-0.0153\\ b_4=0.00150\end{array}$
& $\begin{array}{cc} a_1=0.730\\ a_2=0.372\\ a_3=0.0231\\
    b_1=0.244\\ b_2=0.723\\ b_3=-0.0198\\ b_4=0.00366\end{array}$
\tabularnewline\hline
\end{tabular}
\end{center}
\caption{This table summarises the results for the approximate radiation models corresponding to time dependency $\exp{(-j\omega t)}$. For the opposite convention $\exp{(+j\omega t)}$, complex conjugation of the expressions is needed.}
\label{tab:parameters}
\end{table}

\section*{Acknowledgements}
The study presented in this paper was conducted with the support of the French National Research Agency \textsc{anr} within the \textsc{Consonnes} project. The authors wish to thank Ch.~Morfey and P.~O.~Mattei for useful discussion. \protect{A.N.N.} wishes to express thanks to the Laboratoire de M\'ecanique Physique, Universit\'e Bordeaux~1, for hosting him.


\begin{thebibliography}{10}
\expandafter\ifx\csname url\endcsname\relax
  \def\url#1{\texttt{#1}}\fi
\expandafter\ifx\csname urlprefix\endcsname\relax\def\urlprefix{URL }\fi
\bibitem{levine:1948}
H.~Levine, J.~Schwinger, On the radiation of sound from an unflanged circular
  pipe, Physical Review 73~(4) (1948) 383--406.
\bibitem{nomura:1960}
Y.~Nomura, I.~Yamamura, S.~Inawashiro, On the acoustic radiation from a flanged
  circular pipe, Journal of the Physical Society of Japan 15 (1960) 510--517.
\bibitem{ando:1969}
Y.~Ando, On sound radiation from semi-infinite circular pipe of certain wall
  thickness, Acustica 22~(4) (1969) 219--225.
\bibitem{norris:1989}
A.~N. Norris, I.~C. Sheng, Acoustic radiation from a circular pipe with an
  infinite flange, Journal of Sound and Vibration 135~(1) (1989) 85--93.
\bibitem{dalmont:2001c}
J.-P. Dalmont, C.~J. Nederveen, N.~Joly, Radiation impedance of tubes with
  different flanges : numerical and experimental investigations, Journal of
  Sound and Vibration 244~(3) (2001) 505--534.
\bibitem{davies:1980b}
P.~O. A.~L. Davies, J.~L. Bento~Coelho, M.~Bhattacharya, Reflection
  coefficients for an unflanged pipe with flow, Journal of Sound and Vibration
  72~(4) (1980) 543--546.
\bibitem{causse:1984}
R.~Causs\'e, J.~Kergomard, X.~Lurton, Input impedance of brass musical
  instruments---comparison between experiment and numerical models, Journal of
  the Acoustical Society of America 75~(1) (1984) 241--254.
\bibitem{rienstra:2006}
R.~W. Rienstra, Impedance models in time domain, including the extended
  helmholtz resonator model, in: 12th AIAA/CEAS Aeroacoustics Conference, Vol.
  2006-2686, Cambridge, MA, USA, 2006.
\bibitem{matlab} Matlab$\circledR$, The MathWorks, Inc.
\bibitem{zorumski:1973}
W.~E. Zorumski, Generalized radiation impedances and reflection coefficients of
  circular and annular ducts, Journal of the Acoustical Society of America
  54~(6) (1973) 1667--1673.
\bibitem{lapack}
Lapack Users'Guide, Society for Industrial and Applied Mathematics, ISBN 0-89871-447-8.
\bibitem{mcdaniel:1999}
J.~G. McDaniel, Applications of the causality condition to one-dimensional acoustic reflection problems, Journal of the Acoustical Society of America 105~(5) (1999) 2710--2716.
\bibitem{doutaut:1998}
V. Doutaut, D. Matignon, A. Chaigne, Numerical simulations of xylophones. II. Time-domain modeling of the resonator and of the radiated sound pressure, Journal of the Acoustical Society of America 104~(3) (1998) 1633--1647.
\end{thebibliography}
\end{document}